\documentclass[aps,preprint,showpacs,showkeys,floatfix,longbibliography,superscriptaddress,pra]{revtex4-1}

\usepackage{graphicx}
\usepackage{dcolumn}
\usepackage{amsmath}
\usepackage{amssymb}
\usepackage{epstopdf}
\usepackage{rotating}
\usepackage{xcolor}
\usepackage{soul}
\usepackage{natbib}
\usepackage{url}

\begin{document}

\title{A comparison of normalized gain and Cohen's \textit{d} for analyzing gains on concept inventories}
\keywords{Concept Inventories, Normalized Learning Gain, Effect Size, Student Learning, STEM Education}

\author{Jayson M. Nissen}\affiliation{Department of Science Education, California State University Chico, Chico, CA, 95929, USA} 
\author{Robert M. Talbot}\affiliation{School of Education and Human Development, University of Colorado Denver,  Denver, CO, 80217, USA} 
\author{Amreen Nasim Thompson}\affiliation{School of Education and Human Development, University of Colorado Denver,  Denver, CO, 80217, USA} 
\author{Ben Van Dusen}\affiliation{Department of Science Education, California State University Chico, Chico, CA, 95929, USA} 

\begin{abstract}
Measuring student learning is a complicated but necessary task for understanding the effectiveness of instruction and issues of equity in college STEM courses. Our investigation focused on the implications on claims about student learning that result from choosing between one of two commonly used metrics for analyzing shifts in concept inventories. The metrics are normalized gain (\textbf{g}), which is the most common method used in physics education research and other discipline based education research fields, and Cohen's \emph{d}, which is broadly used in education research and many other fields. Data for the analyses came from the Learning About STEM Student Outcomes (LASSO) database and included test scores from 4,551 students on physics, chemistry, biology, and math concept inventories from 89 courses at 17 institutions from across the United States. We compared the two metrics across all the concept inventories. The results showed that the two metrics lead to different inferences about student learning and equity due to the finding that \textbf{g} is biased in favor of high pretest populations. We discuss recommendations for the analysis and reporting of findings on student learning data.
\keywords{Concept Inventories, Normalized Learning Gain, Effect Size, Student Learning, STEM Education}
\end{abstract}

\maketitle

\section{Introduction}
The methods for measuring change or growth and interpretations of results have been hotly discussed in the research literature for over 50 years \cite{Bereiter1963}. Indeed, the idea of simply measuring a single state (let alone change) in an individual's understanding of a concept, conceptualized as a latent construct, is wrought with issues both philosophical and statistical \citep{Cronbach1970}. Despite these unresolved issues, education researchers use measurement of growth for quantifying the effectiveness of interventions, treatments, and innovations in teaching and learning. Gain scores and change metrics, often referenced against normative or control data, serve as a strong basis for judging the efficacy of innovations. As  researchers commonly measure change and report gains and effects, it is incumbent on researchers to do so in the most accurate and informative manner possible.

\par 	In this work, we collected data at scale and compared it to existing normative data to examine several statistical issues related to characterizing change or gain in student understanding. The focus of our analyses in this investigation are on student scores on science concept inventories (CIs). CIs are research-based instruments that target common student ideas or prior conceptions. These instruments are most often constrained response (multiple choice) and include these common ideas as attractive distractors from the correct response. There exist a multitude of CIs in use across Biology, Chemistry, and Physics (our target disciplines) and in other fields (e.g. Engineering and Math). While CIs are common, the strength of their validity arguments varies widely and some lack normative data. All the CIs used in our work have at least some published research to support their validity and they align with our proposed uses for the scores.

\par A principal tool in quantitative research is comparison, which leads to the frequent need to examine different instruments and contexts. Complications arise in these cross contextual comparisons because the instruments used may have different scales and the scores may greatly vary between populations. For example, some CIs are designed to measure learning across one semester while others are designed to measure learning across several years. Instructors could use both instruments in the same course but  they  would, by design,  give very different results. To compare the changes on the two instruments, researchers need to  standardize the change in the scores, which researchers commonly do by dividing the change by a standardizing coefficient. Unlike in Physics Education Research (PER) and other Discipline Based Education Research (DBER) fields, the social science and education research fields typically use a standardizing coefficient that is a measure of the variance of the scores.

\par The most common gain measurement used in PER and other DBER fields when analyzing CI data is the average normalized gain, \textbf{g}, shown in Equation 1 \cite{Hake1998}. In this equation, the standardizing coefficient is the maximum change that could occur. Hake adopted this standardizing coefficient because it accounted for the smaller shift in means that could occur in courses with higher pretest means. He argued that \textbf{g} was a suitable measure because it was not correlated with pretest mean, whereas posttest mean and absolute gain were correlated with pretest mean and were not suitable measures. He also argued that this normalization allowed for ``a consistent analysis over diverse student populations with widely varying initial knowledge states'' \cite[p.~66]{Hake1998} because the courses with lecture-based instruction \emph{all} had low \textbf{g}, courses with active-engagement instruction primarily had medium \textbf{g}, and no courses had high \textbf{g}. Hake then used this reasoning to define high \textbf{g} (\textbf{g} \textgreater 0.7), medium \textbf{g} (0.7 \textgreater \textbf{g}  \textgreater 0.3), and low \textbf{g} courses (\textbf{g} \textless 0.3).

\begin{equation}
\mathbf{g}= \frac{\bar{x}_{post}-\bar{x}_{pre}}{100\%-\bar{x}_{pre}}
\end{equation}

\par Since Hake published his 1998 paper using \textbf{g}, it has been widely used with the article being cited over 3,800 at the time when this manuscript is being prepared as tracked by Google Scholar. Accordingly, there exists a large amount of gain data expressed in terms of \textbf{g} in the research literature, which serves as normative data for other studies. While the use of \textbf{g} does not align with the practices of the broader social science fields, it would be naive to dismiss this metric as unimportant. However, as noted above and discussed in more detail below, some issues exist with \textbf{g}.

\par The broader field of education research primarily uses the effect size metric as the preferred method for measuring change. The most commonly used effect size metric is Cohen's \emph{d} \citep{Cohen1977}. In effect size metrics, a measure of the variance in the distribution of scores is the standardizing coefficient rather than the maximum possible gain, which \textbf{g} uses. An example of Cohen's \emph{d} is given by Equation 2, where the standardizing coefficient \emph{s} is the pooled standard deviation of the pre- and posttests (discussed further below).

\begin{equation}
\emph{d}= \frac{\bar{x}_{post}-\bar{x}_{pre}}{s}
\end{equation}

\par Researchers have extensively investigated the utility and limitations of \emph{d}, while the research investigating \textbf{g} is limited. In contrast to Hake's (1998) earlier finding, \citet{Coletta2005} found that \textbf{g} was correlated with pretest means. \citet{Willoughby2009} found that inferences based on \textbf{g} suggested gender inequities existed in college STEM courses even though several other measures indicated that there were no gender inequities in those courses. Furthermore, researchers use several different methods for calculating \textbf{g}, which can lead to discrepant findings \citep{Bao2006,Marx2007}. Researchers have also identified issues for \emph{d}. For example, \emph{d} exaggerates the size of effects when measuring changes in small samples \citep{Grissom2012}. Cohen's \emph{d} is based on the t-statistic and the assumptions of normality and homoscedasticity in the test scores used to generate it \cite{Grissom2012}. CI data frequently fails to meet the assumptions of normality and homoscedasticity because of floor and ceiling effects and outliers. We expect that any problems that this creates for \emph{d} are also applicable to \textbf{g}. However, we are not aware of any research on these assumptions pertaining to \textbf{g}.

\section{Purpose}

Both \emph{d} and \textbf{g} have limitations. Our purpose in this investigation was to empirically compare concept inventory gains using both \textbf{g} and \emph{d} to investigate the extent to which they lead to different inferences about student learning. In particular, our concern was that \textbf{g} favors high pretest populations, which leads to skewed measures of student learning and equity. This particularly concerned us because researchers use \textbf{g} as the de facto measure of student learning in PER and other DBER researchers have used it despite there being few investigations of the validity of \textbf{g} and known problems with its efficacy. We compared \textbf{g} to \emph{d} since \emph{d} is gaining use in DBER and is the comparable de facto measure in the much larger fields of sociology, psychology, and education research where researchers have extensively studied its validity, utility, and
limitations. 
\section{Background on Measuring Change}

In this section, we provide a foundation for our motivations and work. First, we discuss the development and use of CIs in undergraduate science education research.  We then discuss statistical issues related to measuring change before reviewing the uses of the average normalized gain in analyzing scores from CIs. Finally, we discuss Cohen's \emph{d} and its use in the context of best practices for presenting data and findings.

\subsection{Rise in the use of CIs to measure student knowledge}

CIs provide ``data for evaluating and comparing the effectiveness of instruction at all levels.'' \citet{Hestenes1992}. They typically consist of banks of multiple-choice items written to assess student understanding of canonical concepts in the sciences, mathematics, and engineering. Researchers generally develop CIs through an iterative process. They identify core concepts with expert feedback and use student interviews to identify common preconceptions and provide wording for distractors. CIs exist for core concepts in most STEM fields, see \cite{Smith2010} for a thorough review and discussion. Though it is unclear exactly how many CI's exist, one of the most widely used CIs, the Force Concept Inventory \citep{Hestenes1992}, has been cited more than 2,900 times at the time when this manuscript is being prepared according to Google Scholar.
 
\par Researchers often use CIs as the outcome measures for evaluative studies to find out if an instructional intervention has an effect on learning relative to a control condition. To facilitate this use, researchers administer CIs pre- and post-instruction, and they compare gains observed in treatment groups to gains observed in a control condition. CIs tend to measure conceptual understanding at a \emph{big picture} level. This means that if students conceive of science learning as a matter of primarily memorizing definitions and formulas (consistent with a more \emph{traditional} conception of teaching and learning), they are unlikely to do well on most CIs. Several studies have used CIs to compare the impact of research-based pedagogies to more traditional pedagogies \citep{Hake1998,Freeman2014,VonKorff2016} and to investigate equity in STEM courses by comparing knowledge and learning of majority and underrepresented minority students \citep{Madsen2013,Brewe2010,Rodriguez2012,Lorenzo2006}. These types of investigations motivate instructors to adopt active learning in courses throughout the STEM disciplines \citep{Singer2012}.
 
\par Because researchers often compare scores for different CIs administered to different populations, they often use a change metric that is standardized and free from the original scale of the measurement. This change metric is often \textbf{g} for DBER studies, but some DBER studies have used \emph{d}. One particular case that focused our current investigation on comparing \textbf{g} and \emph{d}, was the use of \textbf{g} and test means to conclude, ``In the most interactively taught courses, the pre-instruction gender gap was gone by the end of the semester'', \cite[p.~1]{Lorenzo2006}. A finding that \citet{Rodriguez2012} later called into question when their, ``analysis of effect sizes showed gender still impacted FCI scores and that the effect was never completely eliminated'' \cite[p.~6]{Rodriguez2012}.

\subsection{Some issues in measuring change}
Discussions in the measurement literature on quantifying change can be sobering. A classical and often cited work in this area is that of \citet{Cronbach1970}, which raised issues of both the reliability and validity of gain scores. Based on Classical Test Theory, the authors argue that the prime issue of reliability has to do with the systematic relationship between error components of true scores derived from independent, but ``linked'' observations. Consider a common situation in CI use in which the same test is given as both a pre- and posttest. One could argue that the observations (pre and post) are independent measurements since they are taken at different time points, but they are actually \emph{linked} since the measurements are from the same group of students. Because those students had responded to the same instrument at the pretest administration, their posttest scores are likely correlated with their pretest scores due to a shared error component between the two scores. One can correct for this (often overstated) correlation due to the shared error components, but the correction is not always straightforward. \citet{Bereiter1963} calls this the ``over correction under correction dilemma.''  Cronbach and Furby discuss this dilemma at length and offer various methods to dissattenuate the correlation. However, the authors seem to see these correction methods as a workaround for the real issue of linked observations. In their summary discussion, Cronbach and Furby actually state that ``investigators who ask questions regarding gain scores would ordinarily be better advised to frame their questions in other ways'' (p.~80). Despite their persistent statistical issues, gain measurements are widely used in education research due to their great utility. Acknowledging these issues while leveraging the utility requires researchers to be diligent and transparent in their methods and presentation.   

\par Another issue with gain scores has to do with the actual scale of the scores, which Bereiter refers to as the ``physicalism-subjectivism dilemma.'' The issue here is related to the assumption of an interval scale on the construct of interest when using raw scores, or when using gain scores that are normalized on that same scale (as in using \textbf{g}). In other words, the gain metric (\textbf{g}) is scaled in terms of the measure itself (e.g., Newtonian Thinking as measured by a CI) and is assumed to be intervally scaled. A potential solution here is to change the scaling to something that ``seems to conform to some underlying psychological units'', \cite[p.~5]{Bereiter1963}. In this case, the scaling factor (or ``standardization coefficient'') is not based on the scale of the measure (e.g., raw scores on a CI, or Newtonian Thinking) but rather in a \emph{standard} unit such as the variance of the score distribution. In this way, the gain metric is transformed out of the scale of the measure (e.g., Newtonian Thinking) and into a construct independent, standardized scale (e.g., based on variance). Transforming the scale can make cross-scale comparisons possible, and also may highlight potential inequities brought on by remaining in the scale of the measure itself. This latter approach is how the dilemma is addressed when using the effect size metric (discussed further below). For a more detailed discussion of these issues related to measuring change in Classical Test Theory see \citep{Talbot2013}.

\subsection{The Average Normalized Gain}
\citet{Hake1998} developed the average normalized gain (\textbf{g}) as a way to normalize average gain scores in terms of how much gain could have been realized. Hake interpreted \textbf{g} from pre-post testing ``as a rough measure of the effectiveness of a course in promoting conceptual understanding'' (p. 5). His work was seminal in PER and led to the broad use of \textbf{g} throughout DBER. The breadth of its uptake led to at least three different methods for calculating \textbf{g} to be in common use. The original method proposed by Hake calculates \textbf{g} from the group means and is shown in Equation 1. A second method that is more commonly used \citep{VonKorff2016} is to calculate the normalized gain for each individual student $(\mathbf{g}_{I})$ to characterize that student's growth, and to then average the normalized gains for all the individuals to calculate \textbf{g} for the group. \citet{Bao2006} provides an in depth discussion of the affordances of these two methods, but Bao and Hake both state that in almost all cases the two values are within 5\% of one another.
 
\par \citet{Marx2007} proposed a third method, normalized change (\textbf{c}),  in response to several shortcomings of $(\mathbf{g}_{I})$. These shortcomings included: a bias towards low pretest scores, a non-symmetric range of scores ($-\infty$~ to 1), and a value of  -$\infty~$ for any posttest score when the student achieves a perfect pretest score. These limitations inhibit the ability to calculate the average normalized gain for a class by averaging the individual student normalized gains. Instead, they offer a set of rules for determining \textbf{c} based on whether the student gains from pre- to posttest, worsens, or remains at the same score. Their metric results in values of \textbf{c} ranging from -1 to +1. However, \textbf{c} is still sensitive to the distribution of pre and posttest scores in a way that might be ``related to certain features of the population'' \cite{Bao2006}, as it is still normalized on the same scale as the measure itself, an issue raised by \citet{Bereiter1963} and discussed above.
 
\par One particular concern with gain metrics, and with \textbf{g} specifically, has to do with the possibility that these metrics can be biased for or against different groups of students. As \citet{Rodriguez2012} point out, researchers can define equity in several ways. This choice combined with potential bias in the gain metrics leaves open the possibility that results may not represent the actual status of equity in the classroom, for which gain metrics serve as simple but incomplete indicator. For example,  \citet[p.~1]{Willoughby2009} found that ``males had higher learning gains than female students only when the normalized gain measure was utilized. No differences were found with any other measures, including other gain calculations, overall course grades, or individual exams.'' One might expect this to be the case when pretest score is part of the standardization coefficient, since pretest is likely correlated with previous education, and therefore opportunity and even socioeconomic status. Indeed, \citet[p.~1]{Coletta2005} ``found a significant, positive correlation between class average normalized FCI gains and class average preinstruction scores.'' This finding is aligned with \citet{Marx2007} conclusion that \textbf{g} is biased by pretest scores, however, they found it was biased the opposite direction.

\subsection{The Effect Size Metric}

One of the most widely used standardized effect size metrics is Cohen's \emph{d}. Cohen's \emph{d} normalizes (i.e., scales) the difference in scores in terms of the standard deviation of the observed measurements. In essence, it is the difference between Z (standard) scores. This results in a ``pure'' number free from the original scale of measurement \citep{Cohen1977}.  As a result, \emph{d} meets the need for ``\ldots a measure of effect size that places different dependent variable measures on the same scale so that results from studies that use different measures can be compared or combined'', \citet{Grissom2012}.
 
\par 	As a consequence of using the standard deviation, \emph{d} assumes that the populations being compared are normally distributed and have equal variances. Accordingly, the standard deviation used to calculate \emph{d} is that of either sample from the population since they are assumed to be equal. However, in practical applications the pooled standard deviation, Equation 3, of the two samples is used since the standard deviations of the two samples often differ. The pooled standard deviation ($s_{pooled}$) is a weighted average of the standard deviations of the two samples using the size of the samples (\emph{n}) to weight the respective standard deviations. In the case of dependent data such as matched pre- and posttests the sample size (\emph{n}) for both samples is the same and can be factored out of the equation.

\begin{equation}
s_{pooled}=\sqrt{\frac{(n_{1}-1)s_{1}^{2} + (n_{2}-1)s_{2}^{2}}{n_{1}+n_{2}-2}}
\end{equation}

\begin{equation}
d_{dep}= \frac{\bar{x}_{post}-\bar{x}_{pre}}{ \sqrt{s_{pre}^2 + s_{post}^2 -2r s_{pre}  s_{post}}}*\sqrt{2(1-r)}
\end{equation}

\par 	Using either the equal pre- and posttest standard deviations or the pooled standard deviations assumes that the samples (pre- and posttest) are independent and therefore does not take into account or correct for the correlation between measurements made at pre and post (the ``dilemma'' discussed above from Bereiter). The calculation for \emph{d} accounting for the dependence between pre- and posttest \citep{Lakens2013} is shown in Equation 4. Equation 4 is similar to Equation 2 in that it represents the difference in the means divided by the standard deviation (in this case, $s_{pooled}$), noting that there are no sample sizes (\emph{n}'s) in Equation 4 because they factor out of the equation since they are equal. Equation 4 differs in that it includes a correction factor to dissattenuate the effect size based on the correlation (\emph{r}) between the pretests and posttests. When the standard deviations are equal then the dependent and independent forms of Cohen's \emph{d} are equal. If the two are not equal, then the dependent form of Cohen's \emph{d} is always larger because the correlation accounts for some of the variance in the data and thereby reduces the standard deviation. Cohen's \emph{d} can also be calculated from the t-test statistic and this can serve to further elucidate the dependent-independent issue. \citet{Dunlap1996} present an example of calculating a t-statistic between two means when assuming the samples are independent, and again when assuming dependence for the same sample means. When running dependent analyses the ``...correlation between the measures reduces the standard error between the means, making the differences across the conditions more identifiable'' \citep[p.~171]{Dunlap1996}. Thus, taking into account the dependence between the data results in a larger t-statistic because the difference in the means is divided by a smaller standard error.  Cohen's \emph{d} can be directly calculated from the dependent or independent t-statistic. This is why the dependent form of \emph{d} is always larger than the independent form. In practice, many researchers use  the independent form of \emph{d} given in Equation 2 and do not account for correlations between pre- and posttest. For example, \citet{Rodriguez2012} used the dependent samples Cohen's \emph{d} in their reanalysis of earlier pre-post data that did not provide the correlations between pretest and posttest scores. \citet{Dunlap1996} recommend this practice, arguing that  that correlation does not change the size of the difference between the means but only makes the difference more noticeable by reducing the standard error. \citet{Morris2002} agree that using the independent calculation for \emph{d} with dependent data is an acceptable practice so long as all researchers are aware of the issue and any effect sizes being compared are calculated in the same way.

\section{Research Questions}
Given our purpose of comparing \textbf{g} and \textit{d}, our specific research questions were: 
\begin{enumerate}
\item To what extent did the relationships between \textbf{g} and \textit{d} and their relationships to the descriptive statistics used to calculate them indicate that they were biased toward different populations of students? 
\item To what extent did disagreements between \textbf{g} and \textit{d} about the learning for groups of students with different pretest scores confirm any biases identified while investigating the first research question?
\end{enumerate}
Based on previous research we expected differences in the degree to which \emph{d} and \textbf{g} indicated that a phenomenon (e.g., learning gains or equity) was present. We expected the gain characterized by each metric to vary by student population due to differences in pretest scores across populations. This variation across pretest scores motivated our research because it could bias investigations of equity in college STEM learning. We used the second research question to test any biases we identified in a context (gender gaps) that are frequently investigated in the PER literature and to illustrate how bias in the measures used could skew the results of investigations.

\section{Methods} 
To answer these research questions, we used a large data set of student responses to nine different research-based CIs. This data set was large enough to provide useful and reliable comparisons of effect size measures and to represent CI data in general. We processed the data to remove spurious and unreliable data points and used Multiple Imputations (MI) to replace missing data. To simplify our analysis, we first investigated the similarity of the measures resulting from the three ways to calculate Hake's normalized gain, course averages (\textbf{g}), averaged individual gains ($\mathbf{g_{I}}$), and normalized change (\textbf{c}), to determine if they were similar enough we could conduct our further analyses using only one of those approaches. We then made several comparisons of the effect size measures for \textbf{g} and \emph{d} to inform our research questions. We compared the effect size measures to one another and investigated the relationships of the effect size measures with pretest and posttest means and standard deviations to identify any potential biases in the effect size measures. To test any biases we found and to inform the effects of those biases, we compared the effect size measures for subpopulations within each course that have historically different pretest and posttest means. 
\subsection{Data collection and processing}
Our general approach to data collection and processing was to collect the pre and post data with an online platform. We then applied filters to the data to remove pretests or posttests that were spurious. Instead of only analyzing the data from students who provided both a pretest and a posttest, we used Multiple Imputation (MI) to include all the data in the analyses. Online data collection enabled collecting a large dataset, filtering removed spurious and outlier data that was unreliable, and using MI maximized the size of the sample analyzed and the statistical power of the analyses.
\par We used data from the Learning About STEM Student Outcomes (LASSO) platform that was collected as part of a project to assess the impact of Learning Assistants (LAs) on student learning \citep{VanDusen2016,White2016}. LAs are talented undergraduates hired by university and two-year college faculty to help transform courses \citep{Otero2006}. LASSO is a free platform hosted on the LA Alliance website (https://www.learningassistantalliance.org/) and allows faculty (LA-using or not) to easily administer research-based concept inventories as pre- and posttests to their students online. To use LASSO, faculty provide course-level information, select their assessment(s), and upload a list of student names and emails. When faculty launch an assessment, their students receive emails with unique links to complete their tests online at the beginning and end of instruction. Faculty can track students' participation and send reminder emails. As part of completing the instrument, students answer a set of demographic questions.  Faculty can download all of their students' responses and a summary report that includes a plot of their students' pre- and posttest scores and the course's normalized learning gains (\textbf{g}), and effect sizes (Cohen's \emph{d}). 

\par We processed the data from the LASSO database to remove spurious data points and ensure that courses had sufficient data for reliable measurements. We filtered our data with a set of filters similar to those used by \citet{Adams2006} to ensure that the data they used to validate the Colorado Learning Attitudes about Science Survey (CLASS) were reliable. Their filters included number of items completed, duration of online surveys, and a filter question that directed participants to mark a specific answer. In our experience, Adams and colleagues discussion of filtering the data is unique for physics education researchers. Just as \citet{VonKorff2016} found that few researchers explicitly state which \textbf{g} they used, we found that few researchers explicitly address how they filtered their data. For example, authors in several studies \citep{Kost2009,Kost2010,Nissen2016} that used the CLASS made no mention if they did or did not use the filter question to filter their data, nor do they discuss any other filters they may have applied. The lack of discussion of filtering in these three studies is not a unique choice by these authors. Rather, their choice represents the common practices in the physics education literature. 

\par We included courses that had partial data for at least 10 participants to meet the need for a reliable measure of means without excluding small courses from our analyses. We removed spurious and unreliable data at the student and course level if any of the following conditions were met. 
\begin{itemize}
\item A student took less than 5 minutes to complete that test. We reasoned that this was a minimum amount of time required to read and respond to the test questions.
\item A student answered less than 80\% of the questions on that test. We reasoned that these exams did not reflect student's actual knowledge. 
\item A student's absolute gain (posttest mean minus pretest mean) was 2 standard deviations below the mean absolute gain for that test. In these cases, we removed the posttest scores because we reasoned that it was improbable for students to complete a course and unlearn the material to that extent. 
\item A course had greater than 60\% missing data. Low response rates may have indicated abnormal barriers to participating in the data collection that could have influenced the representativeness of the data from those courses.
\end{itemize}

\begin{table}
\caption{Data after each filter was applied.}
\label{tab:1}       
\begin{tabular}{lccccc}
\hline\noalign{\smallskip}
		&~~None~~	&~~Time~~	&Completion	&Gain	&$\geq60\%$~missing \\
\noalign{\smallskip}\hline\noalign{\smallskip}
Courses	&119		&116		&116			&116		&89			\\
Students	&6,041	&5,677	&5,667		&5,667	&4,551 		\\
Pretests	&5,339	&4,922	&4,899		&4,899	&3,842		\\
Posttests	&4,204	&3,693	&3,685		&3,642	&3,335		 \\
Matched	&3,502	&2,973	&2,917		&2,874	&2,626		 \\
\noalign{\smallskip}\hline
\end{tabular}
\end{table}

\par Filter 1, taking less than five minutes, removed 364 students from the data set. Filter 2, completing less than 80\% of the questions, removed 10 students from the data set. Filter 3, a negative absolute gain 2 standard deviations below the mean, removed 0 students but did remove 43 posttests.  Removing the courses with more than 60\% missing data removed 27 courses and 1,116 students from the analysis.

\par To address missing data, we performed multiple imputations (MI) with the Amelia II package in R \cite{Honaker2011}. The most common method for addressing missing data in PER is to use listwise deletion to only analyze the complete cases, discarding data from any student who did not provide both the pretest and posttest. Though, we know of at least one study in PER that used MI \cite{Dou2016}.  We used MI because it has the same basic assumptions of listwise deletion but it reduces the rate of type I error by using all the available information to better account for missing data \citep{Rubin1996}. This leads to much better analytics than traditional methods such as listwise deletion \citep{Allison2002} that, while they ``...have provided relatively simple solutions, they likely have also contributed to biased statistical estimates and misleading or false findings of statistical significance'', \cite[p.~400]{Buhi2008}. Extensive research indicates that in almost all cases MI produces superior results to listwise deletion \cite{Manly2015,Dong2013}.

\par MI addresses missing data by (1) imputing the missing data \textit{m} times to create \textit{m} complete data sets, (2) analyzing each data set independently, and (3) combining the \textit{m} results using standardized methods (Dong and Peng 2013). The purpose of MI is not to produce specific values for missing data but rather to use all the available data to produce valid statistical inferences \cite{Manly2015}. 
\par     Our MI model included variables for CI used, pretest and posttest scores and durations, first time taking the course, and belonging to an underrepresented group for both race/ethnicity and for gender. The data collection platform (LASSO) provided complete data sets for the CI variables and the student demographics. As detailed in Table I, either the pretest score and duration or the posttest score and duration was missing for 42\% of the students. To check if this rate of missing data was exceptional, we identified 23 studies published in the American Journal of Physics or Physical Review that used pre-post tests. Of these 23 studies, 4 reported sufficient information to calculate participation rates \cite{Nissen2016,Kost2009,Kost2010,Cahill2014}. The rate of missing data in these 4 studies varied from 20\% to 51\% with an average of 37\%. The 42\% rate of missing data in this study was within the normal range for PER studies using pre-post tests. 

\par Based on the 42\% rate of missing data we conducted 42 imputations because this is a conservative number that will provide better results than a smaller number of imputations \citep{Bodner2008}. We analyzed all 42 imputed data sets and combined the results by averaging the test statistics (e.g., means, correlations, and regression coefficients) and using Rubin's Rules to combine the standard errors for these test statistics \citep{Schafer1999}. Rubin's Rules combines the standard errors for the analyses of the MI datasets using both the within-imputation variance and the between-imputation variance with a weighting factor for the number of imputations used. For readers interested in further seeking more information on MI, \citet{Schafer1999} and \citet{Manly2015} are useful overviews of MI. All assumptions were satisfactorily met for all analyses. 

\subsection{Investigating the effect size measures}
To identify and investigate differences between the effect size measures, we used correlations and multiple linear regressions (MLR) to investigate the relationships between the effect size measures and the test means and standard deviations. Correlations informed the variables we included in the MLR models. 

\par 	We calculated Cohen's \emph{d} for each course using the independent samples equation, Equation 2. We used this measure because it is the most commonly used in the physics education research literature and because we expected it to have little to no impact on the analyses \cite{Morris2002}, which we discussed in Section III. D.

\par     To test biases in the effect sizes and their effects on CI data, we used the male and female effect size measures in the aggregated data set. We separated these two groups because male students tend to have higher pretest and posttest means on science concept inventories than female students \citep{Madsen2013,Cunningham2015}. Thus, gender provided a straightforward method of forming populations with different test means and standard deviations. Gender also allowed us to frame our analysis in terms of equity of effects. We defined equity as being the case where a course does not increase pre-existing group mean differences. This definition means that for a course to be equitable the effect on the lower pretest group is equal to or larger than the effect on the higher pretest group.  

\par     For this analysis we calculated the effect sizes for males and females separately. For each effect size measure we then calculated the difference between males and females effect sizes, for example $\Delta _{ \mathbf{g}} = \mathbf{g}_{male} - \mathbf{g}_{female}$. If males in a course had a larger effect size than females in the course then that course was inequitable, $\Delta _{ \mathbf{g}} > 0$. This created four categories into which any two effect size measures would locate each course. Two categories for agreement where both effect sizes said it was either equitable or inequitable and two categories for disagreement where one said equity and the other said inequity. If one of the effect size measures was biased and indicated larger effects on higher pretest mean populations than we expected that one type of disagreement would occur more frequently than the other type. To easily identify differences in the number of courses in the disagreement categories and the size of those disagreements, we plotted the data on a scatter plot. We tested the statistical difference in the distributions using a chi square test of independence with categories for each effect size measure and whether they indicated equity or inequity. 

\subsection{Simplifying the analysis}
The multiple methods for calculating normalized gain for a course complicated our purpose of comparing normalized gain and Cohen's \emph{d}. Therefore, we compared normalized gain calculated using each of the three common methods, which are described in Section III. C, for each course. We calculated \textbf{g} using the average pretest and posttest scores for the course. We calculated the course $\mathbf{g_{I}}$ and \textbf{c} by averaging the individual student scores for each course. Correlations between all three measures were all large and statistically significant, as shown in Table 1. The scatter plots for these three measures are shown in Figure 1.  These results indicated that all three measures were very similar. Therefore, we only used the normalized gain calculated using course averages, \textbf{g}, in our subsequent analyses. 
\par The filters we applied to the data likely minimized the differences between $\mathbf{g_{I}}$ and the other two forms of normalized gain. As \citet{Marx2007} point out, $\mathbf{g_{I}}$ is asymmetric. Students with high pretest scores can have very large negative values for $\mathbf{g_{I}}$, as low as approximately -32, but can only have positive values up to 1. We focused on filtering out spurious and unreliable data that would have likely produced many large negative $\mathbf{g_{I}}$ values for individual students and resulted in larger differences between $\mathbf{g_{I}}$ and the other two normalized gain measures. Nonetheless, the notable differences between the three normalized gain metrics all occurred for $\mathbf{g_{I}}$ being much lower than the other two metrics.
\begin{figure*}

  \includegraphics[width=1\textwidth]{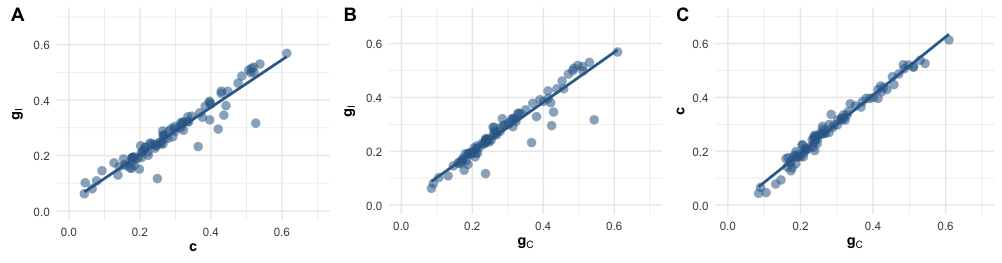}
\caption{Scatter plots comparing the course average value for the three forms of normalized gains.}
\label{fig:scat3g}       
\end{figure*}
%

\begin{table}
\caption{Correlations between the three forms of normalized gain for each course. *** indicates \textit{p}\textless 0.001}
\label{tab:2}       
\begin{tabular}{lcc}
\hline\noalign{\smallskip}
			&\textbf{c}&\textbf{g}\\
\noalign{\smallskip}\hline\noalign{\smallskip}
\textbf{g}		&0.99***	&	 \\
$\mathbf{g_{I}}$&0.93***	&0.93***	\\
\noalign{\smallskip}\hline
\end{tabular}
\end{table}

\section{Findings}
\subsection{Relationship between \textbf{g} and \emph{d}}
Investigating the relationship between \textbf{g} and \emph{d} indicated that there was a large positive relationship between the two measures and it was statistically reliable, r=0.75, \emph{p}\textless0.001. This indicated that \emph{d} and \textbf{g} shared approximately half of their variance in common ($r^{2}$=0.56). Because these two measures serve the same purpose, the 44\% that they \emph{do not} have in common was a large amount. Further investigating the correlations between the effect sizes and their related descriptors, shown in Table 2, revealed large differences between \emph{d} and \textbf{g}.  The correlations between \emph{d} and both pretest mean and pretest standard deviation were small to very small and were not statistically reliable.  In contrast, \textbf{g} was moderately to strongly correlated with both pretest mean and pretest standard deviation. These correlations between \textbf{g} and pretest statistics (0.43 and 0.44 respectively) indicated that approximately one fifth of the variance in normalized gains was accounted for by the score distributions that students had prior to instruction. In contrast, \emph{d} was only weakly associated with both pretest mean and pretest standard deviation. These relationships were strong evidence that \textbf{g} was positively biased in favor of populations with higher pretest scores.

\begin{figure*}
  \includegraphics[width=1\textwidth]{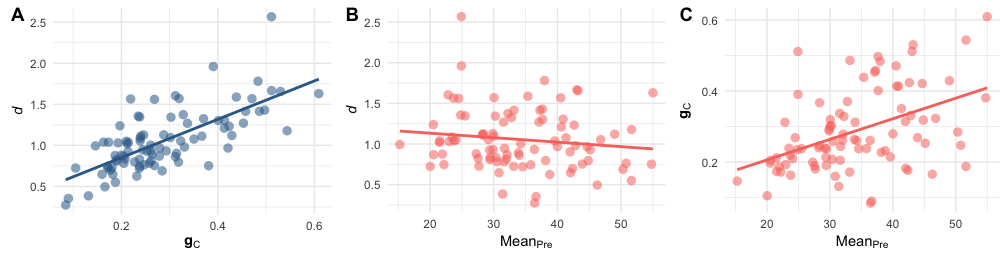}
\caption{Scatter plots for (A) \emph{d} and \textbf{g}, (B)  \emph{d}  and pretest mean, and (C) \textbf{g} and pretest mean.  }
\label{fig:scatdng}       
\end{figure*}
%

\begin{table*}
\caption{Correlations between the effect size measures and test statistics. * \emph{p}\textless0.05. ** \emph{p}\textless0.01. *** \emph{p}\textless0.001.}
\label{tab:3}       
\begin{tabular}{lcccccc}
\hline\noalign{\smallskip}
			&\emph{d} 	&\textbf{g}	&Pre. Mean	&Post. Mean	&Gain	&Pre. S.D.		  \\
\noalign{\smallskip}\hline\noalign{\smallskip}
		&		&			  \\
\textbf{g}		&0.75***		&		&			&			&		&			  \\
Pre. Mean		&-0.11		&0.43***		&		&			&		&			  \\
Post. Mean	&0.43***		&0.87***		&0.81***		&		&		&			  \\
Gain			&0.87***		&0.93***		&0.11		&0.67***		&	&			  \\
Pre. S.D.		&-0.18		&0.44***		&0.67***		&0.65***		&0.24*	&		  \\
Post. S.D.		&-0.24		&0.10		&-0.01		&0.08		&0.15	&0.42***		\\
\noalign{\smallskip}\hline
\end{tabular}
\end{table*}

\par     To inform the size of this bias, we ran several models using MLR with \textbf{g} as the dependent variable and independent variables for \emph{d}, pretest mean, and pretest standard deviation. We used \textbf{g} as the dependent variable because this was consistent with the correlations between \textbf{g} and pretest mean indicating that \textbf{g} was positively biased by pretest means; whereas correlations indicated that \emph{d} was not biased. The linear equation for the final model is given in Equation 5. Our focus in these MLRs was on the additional variance explained by each variable in the models, which we measured using the adjusted $r^{2}$. We did not focus on the coefficients, $\beta$, for each variable.

\begin{equation}
\textbf{g}=\beta_{0} + \beta_{1}*d + \beta_{2}*Mean_{pre} + \beta_{3}*S.D._{pre}
\end{equation}

\par The four models for the MLR are shown in Table III. All the models were statistically significant (\emph{p} \textless0.05). Model 1 only included \emph{d} and shows that \emph{d} and \textbf{g} shared 55\% of the same variance as indicated by the adjusted $r^{2}$ value. Adding either pretest mean or pretest standard deviation to the model markedly increased the explained variance to either 82\% or 89\%, Model 2 and Model 3 respectively. Including all three variables in Model 4 explained 92\% of the variance in \textbf{g}. We interpreted this as indicating that the disagreements between \emph{d} and \textbf{g} were largely explained by the pretest means and standard deviations. Because the pretest means and standard deviations explained such a large proportion of the unexplained variance from Model 1 and the correlations indicated that pretest mean and pretest standard deviation were much more strongly related to \textbf{g} than to \emph{d}, these results indicated that \textbf{g} was biased in favor of groups with higher pretest means.

\begin{table*}
\caption{MLR exploring relationships between \textbf{g}, \emph{d}, and population statistics.}
\label{tab:4}       
\begin{tabular}{lcccccccc}
\hline\noalign{\smallskip}
Model		& \multicolumn{2}{c}{Model 1}		& \multicolumn{2}{c}{Model 2}		&\multicolumn{2}{c}{Model 3}	& \multicolumn{2}{c}{Model 4}\\
\noalign{\smallskip}\hline\noalign{\smallskip}
$r^{2} (\%)$		&\multicolumn{2}{c}{55.6}	&\multicolumn{2}{c}{82.4} &\multicolumn{2}{c}{89.0} &\multicolumn{2}{c}{92.1}  \\
adj. $r^{2} (\%)$	&\multicolumn{2}{c}{55.0} &\multicolumn{2}{c}{82.0}			&\multicolumn{2}{c}{88.7} &\multicolumn{2}{c}{91.8}  \\
\emph{p}		&\multicolumn{2}{c}{\textless0.001}	&\multicolumn{2}{c}{\textless0.001}	&\multicolumn{2}{c}{\textless0.001} &\multicolumn{2}{c}{0.02} \\
&&&&&&&&  \\
D.V.			& $\beta$	&\emph{p}	&$\beta$	&\emph{p}	&$\beta$	&\emph{p}	&$\beta$	&\emph{p}	 \\
\noalign{\smallskip}\hline\noalign{\smallskip}
Intercept		&0.05	&0.05	&-0.21	&\textless0.001	&-0.21	&\textless0.001	&-0.26	&\textless0.001  \\
\emph{d}		&0.22	&\textless0.001	&0.24	&\textless0.001	&0.25	&\textless0.001	&0.25	&\textless0.001  \\
Pre. Mean		&		&		&0.01	&\textless0.001	&		&		&0.01	&\textless0.001  \\
Pre. S.D.		&		&		&		&		&0.01	&\textless0.001	&0.01	&\textless0.001  \\
\noalign{\smallskip}\hline
\end{tabular}
\end{table*}

\subsection{Testing the bias in \textbf{g} using populations with different pretest scores}
\begin{figure*}
  \includegraphics[trim={0 0 0 1.5cm},clip,width=0.9\textwidth]{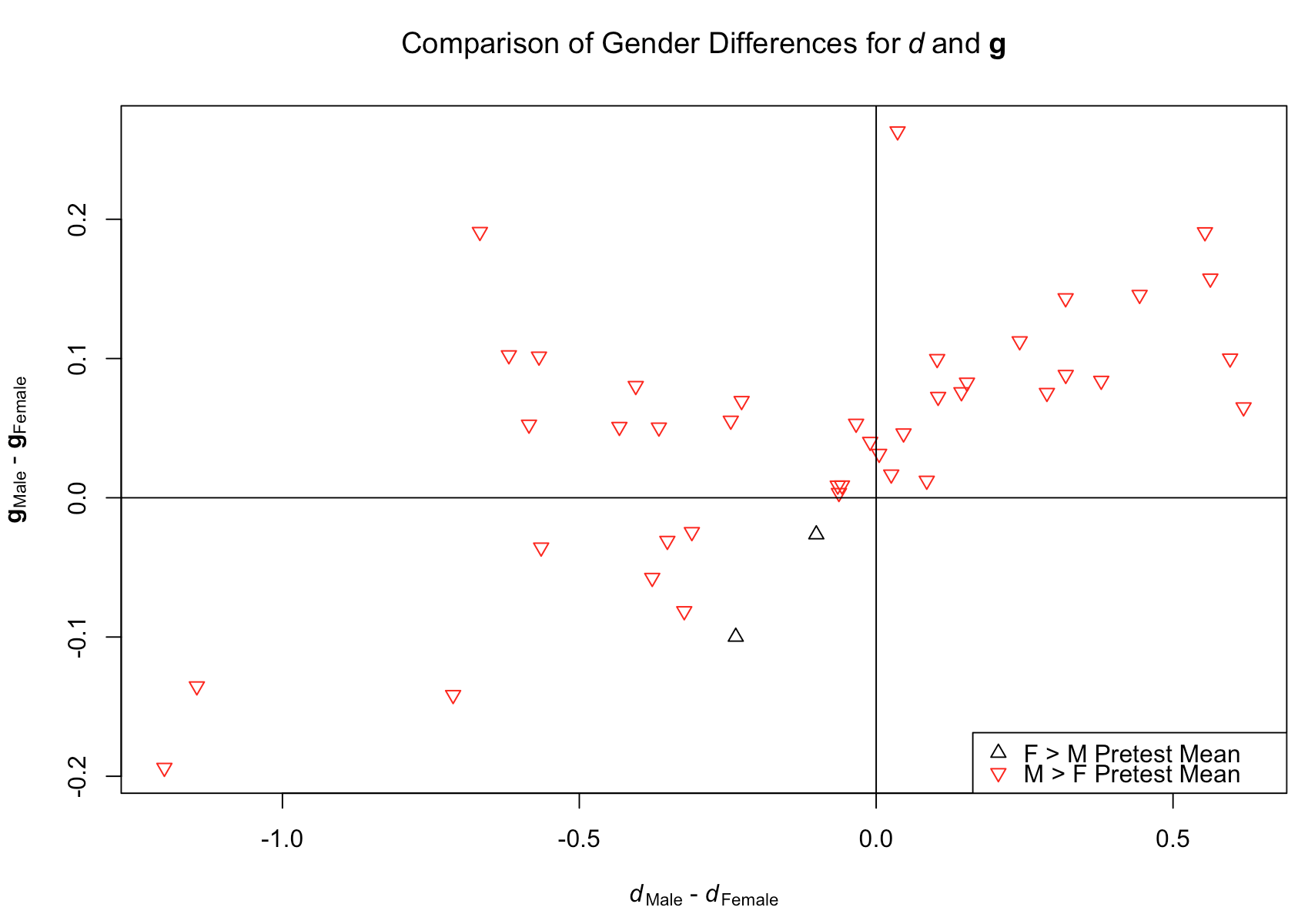}
\caption{Comparison of Gender Differences for \emph{d} and \textbf{g}}
\label{fig:gender}       
\end{figure*}

Results from the MLR Model 2 indicated that a class's pretest mean explained  27\% of the variance in a class's \textbf{g} value that was not explained by \emph{d}. If \textbf{g} is biased in favor of high pretest groups, as the MLR and correlatioms indicated, then we expected the disagreements between \textbf{g} and \emph{d} to skew such that they indicated a bias for \textbf{g} in favor of the high pretest population. To visualize potential bias in \textbf{g} we plotted the difference in \emph{d} on the x axis and the difference in \textbf{g} on the y axis in Figure \ref{fig:gender}. The course marker color shows whether male or female students' pretest means were higher. Almost all of the markers (41 of 43 courses) indicated that male students started with higher pretest means and that the data was consistent with our focus on equity being a larger effect on female students. In total, \textbf{g} showed a larger effect on males in 33 out of 43 courses whereas \emph{d} indicated a larger effect on males in 22 out of 43 courses. Figure \ref{fig:gender} illustrates this bias in \textbf{g} in the difference between Quadrants II and IV. A chi squared test of independences indicated that these differences were statistically reliable, $\chi^{2}(1)=6.10$, \emph{p}=0.013. This difference confirmed that \textbf{g} was biased in favor of the male population, showing that \textbf{g} is biased in favor of populations with higher pretest means. This bias implies that \textbf{g} is not a sufficiently standardized change metric to allow comparisons across populations or instruments with different pretest means and is not a suitable measure of effects.

%
\section{Discussion}
To simplify our comparison of the statistical merits of using \textbf{g} and \emph{d} to measure student learning, we first determined what differences there were between the three methods of calculating \textbf{g}. Our analysis showed that the three methods for calculating normalized gain scores were highly correlated ($r \geq 0.93$). The high level of correlation between the normalized gain values indicated that it made little difference which method we used. This result was encouraging given that many researcher report \textbf{g} scores without discussing which method of calculation they used \citep{VonKorff2016}. The scatter plots (Figure 1) for the three measures of g indicated that the large disagreements between the measures that occurred were cases in which $\mathbf{g_{I}}$ was much lower than both \textbf{g} and \textbf{c}. This discrepancy is consistent with the negative bias in possible $\mathbf{g_{I}}$ scores that led \citet{Marx2007} to develop \textbf{c}. The filters we used to remove unreliable data likely increased the agreement we found between $\mathbf{g_{I}}$ and the other normalized gain measures. However, there were several courses where $\mathbf{g_{I}}$ was noticeably lower than  \textbf{g} and \textbf{c}.  These disagreements indicate two potential problems in the existing literature. Some studies using $\mathbf{g_{I}}$ may have underestimated the learning in the courses they investigated due to the oversized impact of a few large negative $\mathbf{g_{I}}$ values. Alternatively, some studies may have filtered out data with large negative $\mathbf{g_{I}}$ values but not explicitly stated this filtering occurred. Both situations are consistent with Von Korff and colleague's \citep{VonKorff2016} statement that few researchers explicitly state which measure of normalized gain they used. Either situation or a combination of the two make it difficult for researchers to rely on and to replicate the work of those prior studies. 
 
\par     Our comparisons of \textbf{g} and \emph{d} revealed several meaningful differences that indicated that \textbf{g} was biased in favor of high pretest populations. The correlation between \textbf{g} and \emph{d} was strong (r=0.75, \emph{p}\textless 0.001) but was markedly smaller than the correlations between the three different methods of calculating \textbf{g} ($r \geq 0.93$). This correlation of 0.75 meant that \textbf{g} and \emph{d} shared only 56\% of their variance. MLRs indicated that pretest mean and standard deviation explained most of the difference between \textbf{g} and \emph{d}; \textit{d}, pretest mean, and pretest standard deviation accounted for 92\% of the variance in \textbf{g}. Given that \textbf{g} was correlated with these pretest statistics much more strongly than \textit{d}, we concluded that \textbf{g} is biased in favor of populations with high pretest means. We recommend that researchers avoid using all forms of normalized gain and instead report Cohen's \emph{d} and the descriptive statistics used to calculate it, including the correlation between pretest and posttest scores.
 
\par    This bias of \textbf{g} in favor of populations with high pretest means is problematic. The dependence of \textbf{g} on pretest means privileges populations of students who come into a class with more disciplinary knowledge or who perform better on multiple choice exams. This bias disproportionately affects students from traditionally underrepresented backgrounds such as women in physics. When comparing the learning of males and females in our dataset, \textbf{g} identified males as learning more in 33 of 43 courses (77\%) while \emph{d} only identified males as learning more in 23 of 43 courses (53\%), nearly cutting the rate by 1/3 (Figure 1). This difference in measurement indicated that \textbf{g} should not be used for investigations of equity as it overestimated student inequities. Researchers are better served by using statistical methods that analyze individual students posttest scores while controlling for their pretest scores and other variables of interest. All researchers should ensure that they report sufficient descriptive statistics for their work to be included in meta-analyses.
 
\section{Conclusion and Inferences}
The bias in \textbf{g} can harm efforts to improve teaching in college STEM courses by misrepresenting the efficacy of teaching practices across populations of students and across institutions. Students from traditionally underrepresented backgrounds are disproportionately likely to have lower pretest scores, putting them at a disadvantage when instructors make instructional or curricular decisions about an intervention's efficacy based on \textbf{g}. For example, \textbf{g} likely disadvantages instructors who use it to measure learning in courses (e.g., non-major courses) or are at institutions (e.g., 2-year colleges) that serve students who have lower pretest means. This is particularly important for faculty at teaching intensive institutions where evidence of student learning can be an important criterion for tenure and promotion.
 
\par  	Comparing the impact of interventions across settings and outcomes in terms of gain scores requires some form of normalization. Normalized learning gain (\textbf{g}) and Cohen's \emph{d} both employ standardization coefficients to account for the inherent differences in the data. Hake developed \textbf{g} to account for classes with higher pretest means potentially having lower gains due to ceiling effects. By focusing on ceiling effects, \textbf{g} implicitly assumes that any population with a higher pretest score will have more difficulty in making gains than lower pretest populations. This assumption contradicts one of the most well established relationships in education research that prior achievement is a strong predictor of future achievement. Thus, \textbf{g}'s adjustment for potential ceiling effects appears to overcorrect for the problem and results in \textbf{g} being biased in favor of populations with higher pretest means.

\par	Using standard deviation as the standardization coefficient in Cohen's \emph{d} helps to address ceiling effects in that measure. When ceiling effects occur the data compresses near the maximum score. This compression causes the standard deviation to decrease which increases the size of the \emph{d} for the same raw gain. Cohen's \emph{d} also corrects for floor effects by this same mechanism. Instruments that have floor and ceiling effects are not ideal for research because they break the assumption of equal variances on the pre- and posttests and because they are poor measures for high or low achieving students. Instruments designed based on Classical Test Theory, such as the CIs used in this study, mainly consist of items to discriminate between average students and have few items to discriminate between high-performing students or low-performing students. Cohen's \emph{d} may mitigate the limitations of these instruments for measuring the learning of high or low pretest populations of students by accounting for the distribution of tests scores.  When the standard deviation is smaller, as with floor or ceiling effects, the probability of change is lower (i.e., learning is harder) so Cohen's \emph{d} is larger in these cases for the same size change in the means.
 
\par  	In addition to reporting Cohen's \emph{d}, researchers should include descriptive statistics to allow scholars to use their work in subsequent studies and meta-analyses. These descriptive statistics should include means, standard deviations, and sample sizes for each measure used, and correlations between the measures. We include correlations on this list because of the dependent nature of CI pre-post testing is not taken into account by the change metrics we have presented in this paper. As discussed in the background section, this correlation (i.e., \emph{linking}) results in a shared error component that can exaggerate the size of the difference. While it is not a common practice in education research, there are effect sizes and statistical methods that can account for the dependence of pre-post tests in published data when the correlations are reported.
 
\par The bias of \textbf{g} is also an issue for researchers who wish to measure the impact of interventions on student learning. The efficacy of interventions ranging from curricular designs to classroom technologies have been evaluated and scaled-up based on measures of student learning. For these investigations, it is important to have a measure of student learning that is not excessively dependent on the knowledge that students bring to a class. By using the pooled standard deviation, rather than the maximum possible gain as defined by the pretest, as a standardization coefficient, \emph{d} avoids the bias toward higher pretest means while accounting for instrument specific difficulty of improving a raw score. We recommend researchers use \emph{d} rather than \textbf{g} for measuring student learning. Besides being the more reliable statistical method for calculating student learning, the use of \emph{d} by the DBER community would align with the practices of the larger education research community, facilitating more cross-disciplinary conversations and collaborations.

\begin{acknowledgements}
This work was funded in part by NSF-IUSE Grant Numbers DUE-1525338 and DUE-1525115, and is Contribution No. LAA-044 of the International Learning Assistant Alliance. 
\end{acknowledgements}

\bibliography{dandg}   

\end{document}